\RequirePackage{lineno}

\documentclass[twocolumn,superscriptaddress,showpacs,preprintnumbers,amsmath,amssymb]{revtex4}
\usepackage{graphicx}
\usepackage{dcolumn}
\usepackage{bm}
\usepackage{xcolor}
\usepackage{lineno}

\newcommand{\bef}{\begin{figure}}
\newcommand{\eef}{\end{figure}}

\newcommand{\be}{\begin{equation}}
\newcommand{\ee}{\end{equation}}
\newcommand{\bea}{\begin{eqnarray}}
\newcommand{\eea}{\end{eqnarray}}
\begin{document}


\title{$K^{*0}$ production in Au+Au collisions at $\sqrt{s_{\rm NN}}$ = 7.7, 11.5, 14.5, 19.6, 27 and 39~GeV from RHIC beam energy scan}

\affiliation{American University of Cairo, New Cairo 11835, New Cairo, Egypt}
\affiliation{Texas A\&M University, College Station, Texas 77843}
\affiliation{Czech Technical University in Prague, FNSPE, Prague 115 19, Czech Republic}
\affiliation{AGH University of Science and Technology, FPACS, Cracow 30-059, Poland}
\affiliation{Ohio State University, Columbus, Ohio 43210}
\affiliation{University of Kentucky, Lexington, Kentucky 40506-0055}
\affiliation{Panjab University, Chandigarh 160014, India}
\affiliation{Variable Energy Cyclotron Centre, Kolkata 700064, India}
\affiliation{Brookhaven National Laboratory, Upton, New York 11973}
\affiliation{Abilene Christian University, Abilene, Texas   79699}
\affiliation{Instituto de Alta Investigaci\'on, Universidad de Tarapac\'a, Arica 1000000, Chile}
\affiliation{University of California, Riverside, California 92521}
\affiliation{University of Houston, Houston, Texas 77204}
\affiliation{University of Jammu, Jammu 180001, India}
\affiliation{State University of New York, Stony Brook, New York 11794}
\affiliation{Nuclear Physics Institute of the CAS, Rez 250 68, Czech Republic}
\affiliation{Shanghai Institute of Applied Physics, Chinese Academy of Sciences, Shanghai 201800}
\affiliation{Yale University, New Haven, Connecticut 06520}
\affiliation{University of California, Davis, California 95616}
\affiliation{Lawrence Berkeley National Laboratory, Berkeley, California 94720}
\affiliation{University of California, Los Angeles, California 90095}
\affiliation{Indiana University, Bloomington, Indiana 47408}
\affiliation{Warsaw University of Technology, Warsaw 00-661, Poland}
\affiliation{Shandong University, Qingdao, Shandong 266237}
\affiliation{Fudan University, Shanghai, 200433 }
\affiliation{University of Science and Technology of China, Hefei, Anhui 230026}
\affiliation{Tsinghua University, Beijing 100084}
\affiliation{University of California, Berkeley, California 94720}
\affiliation{ELTE E\"otv\"os Lor\'and University, Budapest, Hungary H-1117}
\affiliation{University of Illinois at Chicago, Chicago, Illinois 60607}
\affiliation{University of Heidelberg, Heidelberg 69120, Germany }
\affiliation{Wayne State University, Detroit, Michigan 48201}
\affiliation{Indian Institute of Science Education and Research (IISER), Berhampur 760010 , India}
\affiliation{Kent State University, Kent, Ohio 44242}
\affiliation{Rice University, Houston, Texas 77251}
\affiliation{University of Tsukuba, Tsukuba, Ibaraki 305-8571, Japan}
\affiliation{Lehigh University, Bethlehem, Pennsylvania 18015}
\affiliation{University of Calabria \& INFN-Cosenza, Italy}
\affiliation{National Cheng Kung University, Tainan 70101 }
\affiliation{Purdue University, West Lafayette, Indiana 47907}
\affiliation{Southern Connecticut State University, New Haven, Connecticut 06515}
\affiliation{Central China Normal University, Wuhan, Hubei 430079 }
\affiliation{Technische Universit\"at Darmstadt, Darmstadt 64289, Germany}
\affiliation{Temple University, Philadelphia, Pennsylvania 19122}
\affiliation{Valparaiso University, Valparaiso, Indiana 46383}
\affiliation{Indian Institute of Science Education and Research (IISER) Tirupati, Tirupati 517507, India}
\affiliation{Institute of Modern Physics, Chinese Academy of Sciences, Lanzhou, Gansu 730000 }
\affiliation{Frankfurt Institute for Advanced Studies FIAS, Frankfurt 60438, Germany}
\affiliation{National Institute of Science Education and Research, HBNI, Jatni 752050, India}
\affiliation{University of Texas, Austin, Texas 78712}
\affiliation{Rutgers University, Piscataway, New Jersey 08854}
\affiliation{Institute of Nuclear Physics PAN, Cracow 31-342, Poland}
\affiliation{Max-Planck-Institut f\"ur Physik, Munich 80805, Germany}
\affiliation{Creighton University, Omaha, Nebraska 68178}
\affiliation{Indian Institute Technology, Patna, Bihar 801106, India}
\affiliation{Ball State University, Muncie, Indiana, 47306}
\affiliation{Universidade de S\~ao Paulo, S\~ao Paulo, Brazil 05314-970}
\affiliation{Huzhou University, Huzhou, Zhejiang  313000}
\affiliation{Michigan State University, East Lansing, Michigan 48824}
\affiliation{Argonne National Laboratory, Argonne, Illinois 60439}
\affiliation{United States Naval Academy, Annapolis, Maryland 21402}
\affiliation{South China Normal University, Guangzhou, Guangdong 510631}

\author{M.~S.~Abdallah}\affiliation{American University of Cairo, New Cairo 11835, New Cairo, Egypt}
\author{B.~E.~Aboona}\affiliation{Texas A\&M University, College Station, Texas 77843}
\author{J.~Adam}\affiliation{Czech Technical University in Prague, FNSPE, Prague 115 19, Czech Republic}
\author{L.~Adamczyk}\affiliation{AGH University of Science and Technology, FPACS, Cracow 30-059, Poland}
\author{J.~R.~Adams}\affiliation{Ohio State University, Columbus, Ohio 43210}
\author{J.~K.~Adkins}\affiliation{University of Kentucky, Lexington, Kentucky 40506-0055}
\author{I.~Aggarwal}\affiliation{Panjab University, Chandigarh 160014, India}
\author{M.~M.~Aggarwal}\affiliation{Panjab University, Chandigarh 160014, India}
\author{Z.~Ahammed}\affiliation{Variable Energy Cyclotron Centre, Kolkata 700064, India}
\author{D.~M.~Anderson}\affiliation{Texas A\&M University, College Station, Texas 77843}
% ...

\collaboration{STAR Collaboration}\noaffiliation
\affiliation{}
\date{\today}
\begin{abstract}
We report the measurement of $K^{*0}$ meson at 
midrapidity ($|y|<$ 1.0) in Au+Au collisions at 
$\sqrt{s_{\rm NN}}$~=~7.7, 11.5, 14.5, 19.6, 27 
and 39 GeV collected by the STAR experiment during the RHIC 
beam energy scan (BES) program. The transverse 
momentum spectra, yield, and average transverse 
momentum of $K^{*0}$  are presented as functions of collision centrality and beam energy. The 
$K^{*0}/K$ yield ratios are presented for 
different collision centrality intervals and 
beam energies. The $K^{*0}/K$ ratio in 
heavy-ion collisions are observed to be smaller 
than that in small system  
collisions (e+e and p+p). The $K^{*0}/K$ ratio follows a 
similar centrality dependence to that observed in 
previous RHIC and LHC measurements. The data 
favor the scenario of the dominance of hadronic 
rescattering over regeneration for $K^{*0}$ production in the hadronic phase of the medium. 

\end{abstract}
\maketitle

\section{Introduction}
Resonances are very short-lived particles and provide an excellent probe of properties of QCD medium in heavy-ion collisions (HIC)~\cite{Brown:1991kk}. They decay through strong interactions within roughly $10^{-23}$ seconds or a few fm/c which is of a similar order to the lifetime of the medium created in heavy-ion collisions. Due to their short lifetime, some resonances decay within the medium. Hence, they are subjected to in-medium interactions. During the evolution of HIC, the chemical (CFO) and kinetic (KFO) freeze-out temperatures play important roles. At CFO, the inelastic interactions among the constituents are expected to  cease~\cite{Rapp:2000gy,Song:1996ik,Rafelski:2001hp,Becattini:2005xt,Cleymans:2004bf,Andronic:2005yp}. Afterward, the constituents can interact among themselves via elastic (or pseudo-elastic) interactions until the KFO, when their mean free path increases and all interactions cease. Between CFO and KFO, there can be two competing effects, rescattering and regeneration. The momentum of resonance daughters (e.g pions and kaons from $K^{*0}$) can be altered due to the scattering with other hadrons present in the medium. Thus the parent resonance (e.g. $K^{*0}$) is not reconstructible using the re-scattered daughters. This may result in a reduced resonance yield. On the other hand, resonances may be regenerated via pseudo-elastic interactions (e.g. $\pi K \leftrightarrow K^{*0}$) until KFO is reached. Such regeneration may result in an increase of resonance yield. The $K^{*0}$ regeneration depends on the kaon-pion interaction cross section ($\sigma_{K\pi}$), the time scale allowed for this re-generation, and the medium density. The rescattering depends on resonance lifetime, daughter particle's interaction cross-section with the medium (e.g. $\sigma_{K\pi, \; \pi\pi, \; KK}$), the medium density, and the time scale between CFO and KFO. The final resonance (e.g. $K^{*0}$) yield is affected by the relative strength of these two competing processes.  Since the $\sigma_{\pi\pi}$ is about a factor of five larger than $\sigma_{K\pi}$~\cite{Protopopescu:1973sh,Matison:1974sm,Bleicher:1999xi}, one naively expects a loss of $K^{*0}$ signal due to rescattering over regeneration. Furthermore, the mass peak position and width of resonances may be modified due to in-medium effects and late stage rescattering.


Due to the short lifetime of about 4.16 fm/c, the $K^{*0}$ meson is one of the ideal candidates to probe the hadronic phase of the medium between CFO and KFO. If rescattering plays a dominant role, then one naively expects a smaller resonance to non-resonance particle yield ratio (e.g. $K^{*0}/K$) in central collisions compared to that in peripheral and small system (p+p) collisions. On the contrary, if regeneration is dominant, the above ratio is expected to be larger in central compared to peripheral (and small system) collisions. In previous RHIC~\cite{STAR:2002npn,STAR:2004bgh,STAR:2008twt,STAR:2010avo,PHENIX:2014kia}, SPS~\cite{NA49:2011bfu,NA61SHINE:2020czr}, and   LHC~\cite{ALICE:2014jbq,ALICE:2017ban,ALICE:2019xyr,ALICE:2012pjb,ALICE:2019hyb,ALICE:2019etb,ALICE:2021ptz} measurements, it is observed that the $K^{*0}/K$ ratio is indeed smaller in central heavy-ion collisions than in peripheral, and elementary (e.g. p+p) collisions. The observation indicates the dominance of hadronic rescattering over regeneration. Such an observation is also supported by several transport model calculations~\cite{Bleicher:2002dm,Knospe:2015nva,Singha:2015fia}. The measurement of $K^{*0}$ in the Beam Energy Scan range can provide information on the interactions in the hadronic phase of the medium at these energies.

In this article, we report on the measurement of $K^{*0}$ mesons at midrapidity ($|y|<$ 1.0) using data from Au+Au collisions at $\sqrt{s_{\rm NN}}$ = 7.7, 11.5, 14.5, 19.6, 27 and 39 GeV collected by the STAR experiment during 2010-2014 in the $1^{st}$ phase of the Beam Energy Scan (called BES-I) program. The paper is organized as follows: Section II briefly describes the sub-detectors of STAR used in this analysis along with the event and track selection criteria and the data-analysis methods. The results for $K^{*0}$ mesons, which include transverse momentum ($p_{\rm T}$) spectra, yield ($dN/dy$), average transverse momentum ($\langle p_{\rm T} \rangle$) and ratios to non-resonances are discussed in section III. The results are summarized in Section IV.

\section{Experimental  details and data analysis}
\subsection{STAR detector}
The details of the STAR detector system are 
discussed in~\cite{STAR:2002eio}. The detector configuration during 2010 and 2011 are similar, while during 2014 the Heavy Flavor Tracker~\cite{Contin:2017mck} was installed inside the TPC. Minimum-bias events are selected using the scintillator-based Beam Beam Counter (BBC) detectors. The BBCs are located on the two sides of the beam pipe in the pseudo-rapidity range $3.3 < |\eta| < 5.0$. The Time Projection Chamber (TPC)~\cite{Anderson:2003ur} is the main tracking detector in STAR and is used for track reconstruction for the decay daughters of $K^{*0}$. The TPC has an acceptance of $\pm$ 1.0 in pseudo-rapidity and $2\pi$ in azimuth. With the TPC, one can identify particles in the low momentum range by utilizing energy loss ($dE/dx$) and momentum information. The Time of Flight (TOF)~\cite{Llope:2003ti,Bonner:2003bv} detector can be used to identify particles in the momentum region where the TPC $dE/dx$ bands for pions and kaons overlap. The TOF works on the principle of Multigap Resistive
Plate Chamber (MRPC) technology and provides pseudorapidity coverage $|\eta| < 0.9$ with full $2\pi$ azimuth.

\subsection{Event selection}
Minimum-bias events are selected using the coincidence between the BBC detectors~\cite{Bieser:2002ah}. The primary vertex of each event is reconstructed by finding the best common point from which most of the primary tracks originate. The vertex position along the beam direction ($V_{z}$) is required to be within $\pm$~50 cm for $\sqrt{s_{\rm NN}} \geq $ 11.5 GeV and $\pm$~70 cm for 7.7 GeV in a coordinate system whose origin is at the center of TPC. The vertex in radial direction ($V_{r} = \sqrt{V_{x}^{2} + V_{y}^{2}}$) is required to be smaller than 2.0 cm for all energies except 14.5 GeV where the vertex is not centered at (0, 0) in the $xy$ plane and slightly offset at (0.0, -0.89). Hence the $V_{r}$ is selected  to be $V_{r} = \sqrt{V_{x}^{2} + (V_{y}+0.89)^{2}} < 1$ cm for 14.5 GeV~\cite{STAR:2019vcp}. The $V_{r}$ selection excludes events where the incoming Au nuclei collide with the beam pipe. The above vertex selection criteria also ensure uniform acceptance within the $\eta$ range ($|\eta| < 1.0$) studied. A typical vertex resolution 350 $\mu m$ can be achieved using about 1000 tracks with a maximum 45 hit points in TPC ~\cite{STAR:2013ayu}. The number of good events selected after these criteria are listed in Table~\ref{table1_event}.

\begin{table}
\begin{center}
\caption{Au+Au collision datasets,  vertex position $V_{z}$ and $V_{r}$ selection, number of events analyzed.}
\label{table1_event}
\scalebox{1.0}{
\begin{tabular}{ccccc}
\hline 
\hline
Year & Energy  & $|V_{Z}|$ (cm) &   $V_{r}$ (cm) & Events (M) \\
\hline \hline
\\
2010 & 7.7 GeV & $<$ 70 &   $<$ 2 &  4.7 \\
2010 & 11.5 GeV & $<$ 50 &   $<$ 2 &  12.1 \\
2014 & 14.5 GeV & $<$ 50 &   $<$ 1 &  15.3 \\
2011 & 19.6 GeV & $<$ 50 &   $<$ 2 &  27.7 \\
2011 & 27 GeV & $<$ 50 &   $<$ 2 &  53.7 \\
2010 & 39 GeV & $<$ 50 &   $<$ 2 &  128.5 \\
\\

\hline \hline
\end{tabular}
}
\end{center}

\end{table}

\subsection{Centrality selection}
The collision centrality is determined via a fit to the charged particle distribution within $|\eta| < $ 0.5 in the TPC using a Glauber Monte Carlo simulation~\cite{Miller:2007ri}. The minimum bias triggered events are divided into nine different intervals as 0 -- 5\%, 5 -- 10\%, 10 -- 20\%, 20 -- 30\%, 30 -- 40\%, 40 -- 50\%, 50 -- 60\%, 60 -- 70\% and 70 -- 80\%. The average number of participant nucleons $\langle N_{\mathrm{part}} \rangle$ for BES-I energies are evaluated using a Glauber simulation and are reported in~\cite{STAR:2017sal,STAR:2019vcp}.

\subsection{Track selection}
Good quality tracks are selected by requiring at least 15 hit points
in the TPC. In order to reduce track splitting, the tracks are required to include more than 55\% of the maximum number of hits possible for their geometry. Particles are required to have transverse momentum greater than 0.15 GeV/$c$. To reduce contamination from secondary particles (e.g. weak decay contributions), the distance of closest approach (DCA) to the primary vertex is required to be smaller than 2 cm. Lastly, to ensure uniform acceptance, tracks are required to fall within $\pm 1$ in pseudo-rapidity.

\begin{figure*}[!ht]
\includegraphics[scale=0.65]{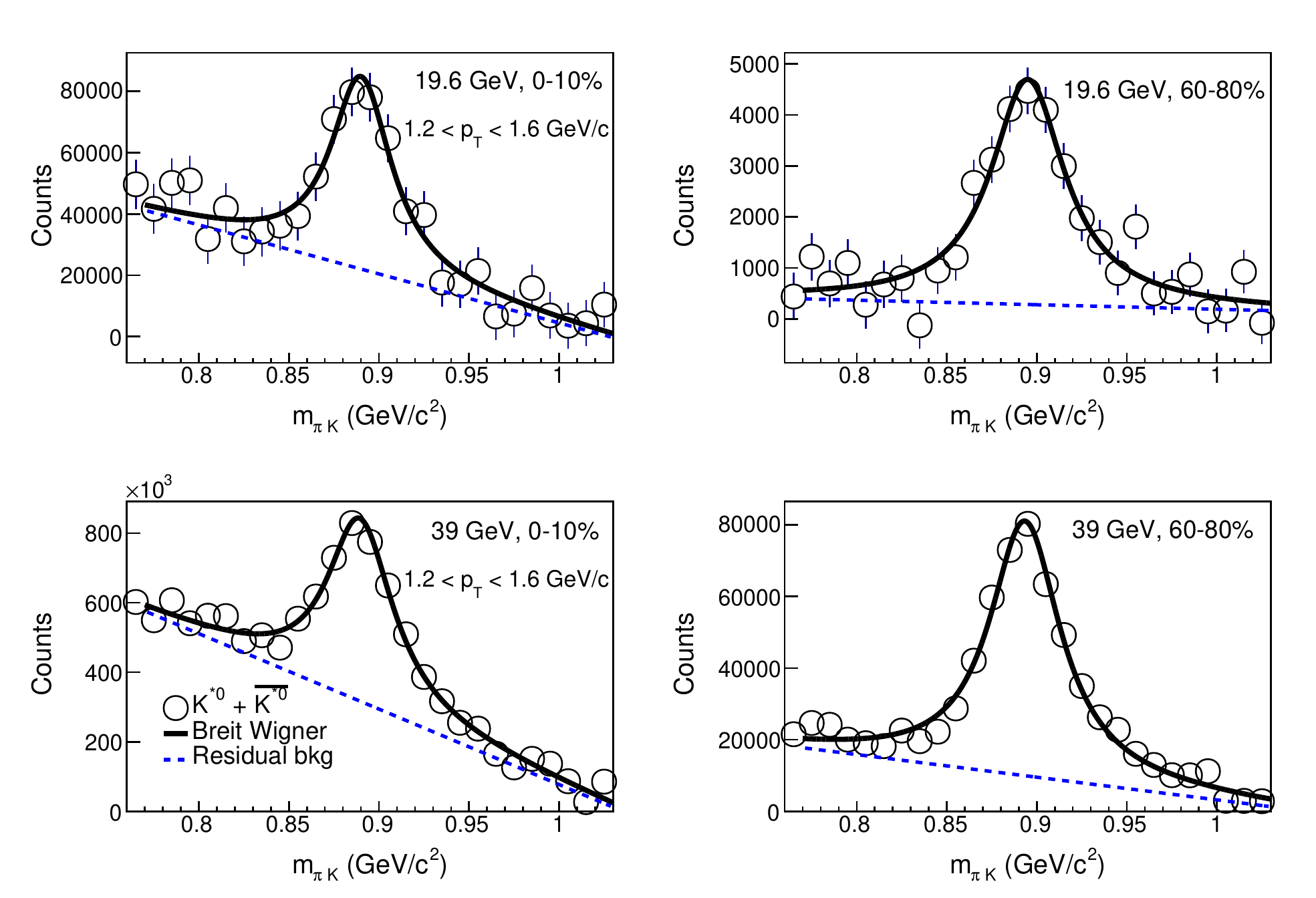}

\caption{The track-rotation combinatorial background subtracted $K\pi$ invariant mass distribution for the 1.2 $< p_{\rm T} <$ 1.6 GeV/$c$ (14.5 and 39 GeV). The data are fitted with a Breit-Wigner function plus a first-order polynomial (as given in equation ~\ref{eqn_bw}) by the solid line. The dashed line represents the residual background only. The uncertainties on the data points are statistical only and shown by bars.}
\label{fig:signal_kstar}
\end{figure*}

\subsection{Particle identification}
Particle identification (PID) is carried out utilizing both the TPC and TOF detectors. The pion and kaon candidates are identified using the energy loss $dE/dx$ of the particles inside the TPC. In the STAR TPC, pions and kaons can be distinguished up to about $0.7$  GeV/$c$ in momenta, while (anti-) protons can be distinguished up to about $1.1$ GeV/$c$ in momenta. Particle tracks in the TPC are characterized by the $N\sigma$ variable, which is defined as:

\begin{equation}
N\sigma (\pi,K) = \frac{1}{R} \mathrm{log} \frac{(dE/dx)_{\mathrm{ meas.}}}{\langle dE/dx \rangle_{\mathrm{theo.}}}, \hspace{-.5em}
\label{eqn-dedx}
\end{equation}

where the $(dE/dx)_{\mathrm{meas.}}$ is the measured energy loss inside the TPC for a track, $\langle dE/dx \rangle_{\mathrm{theo.}}$ is the expected mean energy loss from a parameterized Bichsel function~\cite{Bichsel:2006cs}, and $R$ is the $dE/dx$ resolution which is about 8.1\%. 
The $N\sigma$ distribution is nearly Gaussian at a given momentum and calibrated to be centered at zero for each particle species with a width of unity~\cite{Shao:2005iu}.

The TOF detector extends the particle identification capabilities to intermediate and high $p_{\rm T}$. The TOF system consists of TOF trays and Vertex Position Detectors (VPDs). 
By measuring the time of flight of each particle, we can calculate mass-squared ($m^{2}$) of the corresponding track,
\begin{equation}
m^{2} = p^{2}((t_{\mathrm{TOF}} \times c/l)^{2} -1 ), \hspace{-.5em}
\label{eqn-mass2-tof}
\end{equation}
where $p$ is the momentum, $t_{\mathrm{TOF}}$ is the time of flight, $c$ is the speed of light in vacuum and $l$ is the flight path
length of the particle. The time resolution of TOF is about $\approx$ 80 -- 100 ps. Using the information from the TOF, pions and kaons can be separated up to $p$ $\approx$~1.6 GeV/c, and protons and kaons up to $p$ $\approx$~3.0 GeV/c~\cite{Shao:2005iu}. If the TOF-information is available, $-0.2 < m^{2} < 0.15$ (GeV/c$^{2}$)$^{2}$  and $0.16 < m^{2} < 0.36$ (GeV/c$^{2}$)$^{2}$ is required for selecting pions and kaons respectively. Otherwise we use the TPC $|N\sigma (\pi/K)| < 2.0$ to select pions or kaons.

\subsection{$K^{*0}$ reconstruction}
The $K^{*0}$ (and its antiparticle $\overline{K}^{*0}$) is reconstructed from its hadronic decay channel $K^{*0} (\overline{K}^{*0}) \rightarrow
\pi^{-}K^{+} (\pi^{+}K^{-}) $ (branching ratio $66\%$)~\cite{ParticleDataGroup:2020ssz}. The measurements are performed with the same collision centrality intervals (10\%) for all energies except for $\sqrt{s_{\rm NN}}$ = 7.7 GeV, where the intervals are changed from 10\% to 20\% due to the low charged particle multiplicity at this energy.
The analysis is done by combining both $K^{*0}$ and $\overline{K}^{*0}$, which in the text is denoted by $K^{*0}$, unless specified.

In a typical event, it is impossible to distinguish the decay daughters of $K^{*0}$ from other primary tracks. First, the invariant mass is reconstructed from the unlike sign K$\pi$ pairs in an event (called same-event pairs). The resultant invariant mass distribution contains true  $K^{*0}$ signal and a large random combinatorial background. Due to the large combinatorial background, the $K^{*0}$ invariant mass peak is not  visible. The typical signal to background ratio is within the range 0.002 - 0.02. Hence, the background must be subtracted from the same event distribution. The random combinatorial background is estimated using the daughter track rotation technique. In this analysis, the azimuthal angle of kaon track is rotated by 180\textdegree in a plane normal to particle's momentum vector, which breaks the correlation among the pairs originating from same parent particle. The $K^{*0}$ invariant mass peak is obtained after subtracting the invariant mass distribution of the rotated tracks from the same event invariant mass distribution. The signal peak is observed on top of a residual background. The significance of $K^{*0}$ signal is within the range 5-80 for all beam energies and centralities. It has been observed that the residual background may originate from correlated real $K\pi$ pairs from particle decays, correlated pairs from jets, or correlated mis-identified pairs~\cite{STAR:2004bgh}. 


Figure~\ref{fig:signal_kstar} presents the $K^{*0}$ invariant mass signal in the range 1.2 $< p_{\rm T} <$ 1.6 GeV/c for two beam energies, $\sqrt{s_{\rm NN}}$ = 14.5 and 39 GeV, and for two centralities, 0-10\%  and 60-80\%. The $K^{*0}$ invariant mass distribution is obtained in different transverse momentum bins for different collision centrality intervals for six colliding beam energies. It is fitted with a Breit-Wigner and a first order polynomial function and is defined by,
\begin{equation} \label{eqn_bw}
\begin{split}
\frac{dN}{dm_{\pi K}} = \frac{Y}{2\pi} \times \frac{\Gamma_0}{(m_{\pi K}
  - M_0)^{2} + \frac{\Gamma_{0}^{2}}{4}} \\ + \; ( Am_{\pi
  K} + B), \hspace{-.5em}
\end{split}
\end{equation}
 
The Breit-Wigner function describes the signal distribution while the first order polynomial is included to account for the residual background. Here Y is the area under the Breit-Wigner function; $M_{0}$ and $\Gamma_{0}$ are the mass and width of $K^{*0}$. The $K^{*0}$ invariant mass distribution is fitted within 0.77 $< m_{\pi K} <$  1.04 GeV/$c^{2}$. The invariant mass peak and width of $K^{*0}$ are found to be consistent within uncertainty with previously published STAR measurements in Au+Au and p+p collisions (not shown here) at $\sqrt{s_{\mathrm NN}}$ = 200 GeV~\cite{STAR:2002npn,STAR:2004bgh,STAR:2008twt,STAR:2010avo}. Since the mass and width are consistent between heavy-ion and p+p collisions, it indicate that the $K^{*0}$ line shape may not offer sensitivity to in-medium interactions and rescattering. Since the $K^{*0}$ width is consistent with PDG value within uncertainty, the yield is calculated by keeping the width fixed to the vacuum value to avoid any statistical fluctuation.
The boundary of the fitting range is varied within 0.01-0.02 GeV/$c^{2}$. The resulting variation in the $K^{*0}$ yield is incorporated into the systematic uncertainties. The variation in residual background functions (first and second order polynomials) is also included in the systematic uncertainties. The yield of the $K^{*0}$ is extracted in each $p_{\rm T}$ and collision centrality interval by integrating the background subtracted invariant mass distribution in the range of 0.77  $< m_{\pi K} <$  1.04 GeV/$c^{2}$, subtracting the integral of the residual background function in the same range, and correcting the result to account for the yield
outside this region by using the fitted Breit-Wigner function. This correction is about $\approx$ 10\% of the $K^{*0}$ yield. Alternatively, the yield is extracted by integrating the fitted Breit-Wigner function only. The difference in the measured yield from various yield extraction method is about 5$\%$. As a consistency check, the combinatorial background is also estimated from a mixed event technique. The resultant yield of $K^{*0}$ after the background subtraction is found to be consistent with that from the track rotation method within uncertainties.

\subsection{Detector acceptance and reconstruction and PID efficiency correction}

\begin{figure*}[!ht]
\begin{center}
\includegraphics[scale=0.7]{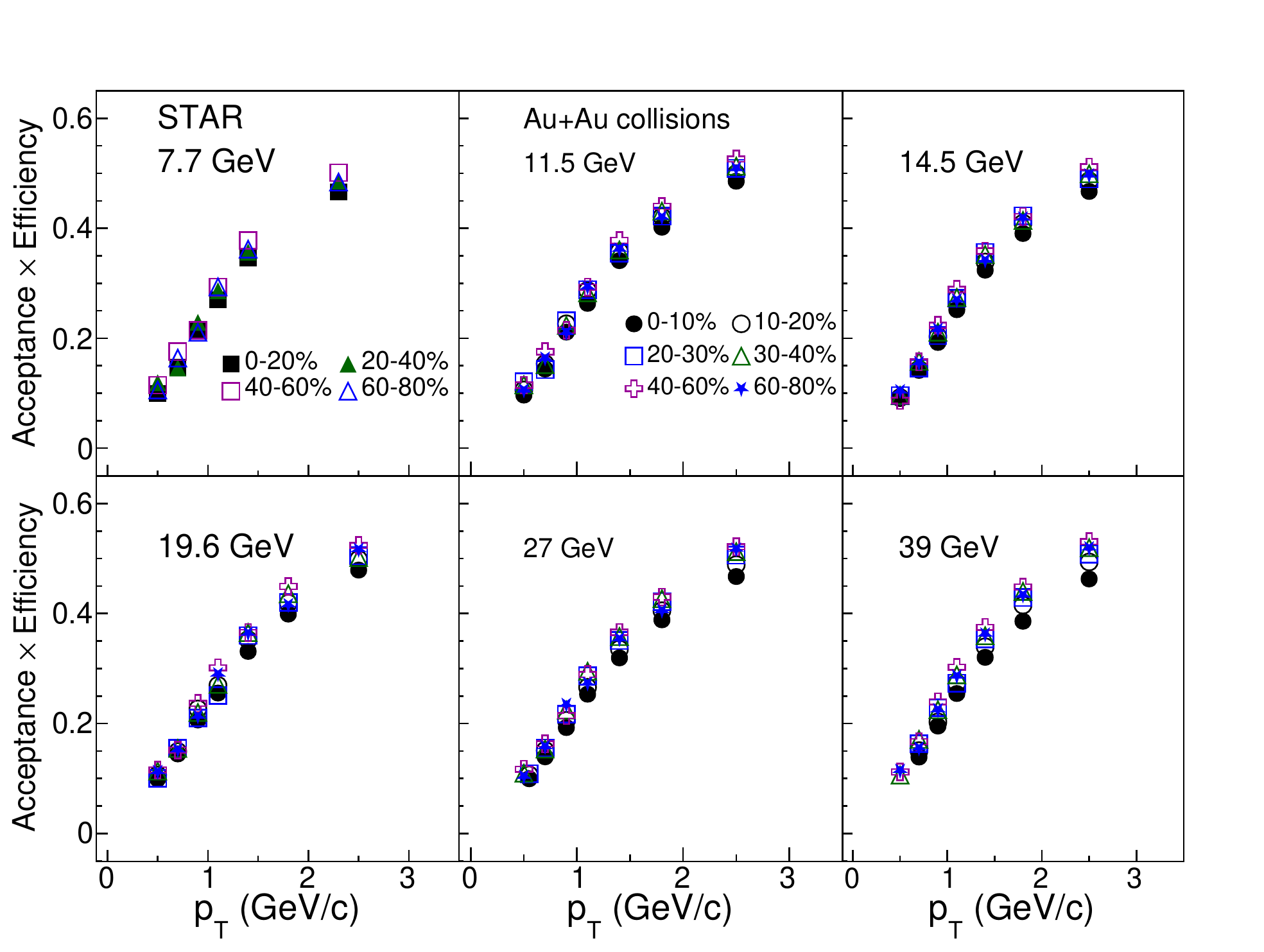}

\caption{The detector efficiency $\times$ acceptance in reconstructing the $K^{*0}$ at various collision centralities in Au+Au collisions at $\sqrt{s_{NN}}$ = 7.7, 11.5, 14.5, 19.6, 27 and 39 GeV. The statistical uncertainties are within the marker size.}
\label{efficiency}
\end{center}
\end{figure*}

The detector acceptance and the reconstruction efficiency ($\epsilon_{\mathrm{acc \times rec}} $) is calculated by using the STAR embedding method. In this process, first $K^{*0}$ is generated with uniform rapidity ($|y|<1.0$), $p_{\rm T}$ ( $0 < p_{\rm T} < 10$ GeV/c) and $\phi$ ($0< \phi < 2\pi$) distribution. The number of $K^{*0}$s generated is about 5\% of the total multiplicity of the event. Then the $K^{*0}$ is decayed and its daughters are passed through the STAR detector simulation in GEANT3 and the TPC Response Simulator~\cite{STAR-GEANT}. The simulated electronic signals are then combined with real data signals to produce a "combined event". This combined event is then passed through the standard STAR reconstruction chain. The reconstruction efficiency $\times$ acceptance ($\epsilon_{\mathrm{acc \times rec}}$) is the ratio of the  number of reconstructed $K^{*0}$s after passing through detector simulation with the same event/track selection parameters used in real data analysis to the input simulated number of $K^{*0}$s within the same rapidity ($|y|<1.0$) interval. Figure~\ref{efficiency} presents the detector acceptance and reconstruction efficiency as a function of $p_{\rm T}$ for different collision centrality intervals in $\sqrt{s_{\rm NN}}$ = 7.7, 11.5, 14.5, 19.6, 27 and 39 GeV collisions. The absence of clear centrality dependence in $\epsilon_{\mathrm{acc \times rec}}$ could be due to the small variation in total multiplicity across the collision centrality and beam energy studied. 

The particle identification efficiency ($\epsilon_{\mathrm{PID}}$) accounts for loss of particles due to TPC $N\sigma$ and TOF mass-squared cuts on $K^{*0}$ daughters. The $\epsilon_{\mathrm{PID}}$ is the product of efficiencies for each decay daughters. The PID efficiency is calculated using the $N\sigma$ and mass-squared distributions in real data. When the $N\sigma$ cuts are applied on pions and kaons, $\epsilon_{\mathrm{PID}}$ for TPC is about 91.1\% and for TOF it is more than 95\%.


\subsection{Systematic uncertainty}
The systematic uncertainties are evaluated bin-by-bin for $p_{\rm T}$ spectra, yield and $\langle p_{\rm T} \rangle$ of $K^{*0}$. The sources of systematic uncertainties in the measurement are (i) signal extraction, (ii) yield extraction, (iii) event and track selections, (iv) particle identification and (v) global tracking efficiency. The systematic uncertainties due to signal extraction are assessed by varying the invariant mass fit range, residual background function (1$^{st}$ order versus 2$^{nd}$ order polynomial) and the invariant mass fit function (non-relativistic versus p-wave relativistic Breit-Wigner function~\cite{STAR:2004bgh}). The systematic in yield calculation is obtained by using histogram integration versus functional integration of the invariant mass distributions. Furthermore, the yield is calculated by keeping the width as a free parameter and fixed to the vacuum value. The variation in the yields are incorporated into the systematic uncertainties. The bounds of event, track quality, and particle identification selection cuts are varied by $\approx$ 10--20\% (e.g. $V_{z}$ selection variation; number of hits in TPC, $|\rm DCA|$, $|N\sigma|$ and TOF-mass$^{2}$ variations), and the resulting difference is included into systematic uncertainties. The uncertainty due to global tracking efficiency is estimated to be 5\% for charged particles~\cite{STAR:2017sal}, which results in 7.1\% for track pairs for $K^{*0}$. The systematic uncertainty in $dN/dy$ and $\langle p_{\rm T} \rangle$ due to the low $p_{\rm T}$ extrapolations are obtained by using different fit functions ($p_{\rm T}$ and $m_{\rm T}$ exponential, and Boltzmann~\cite{STAR:2017sal}) compared to the default Tsallis fit function~\cite{Tsallis:1987eu}. The systematic uncertainties for each of the above sources are calculated as (maximum - minimum)/$\sqrt{12}$ assuming uniform probability distributions between the maximum and minimum values. 
The final systematic uncertainty is the quadratic sum of the systematic uncertainties for each of the above sources ((i)-(v)). The typical average systematic uncertainties in $p_{\rm T}$ spectra, $dN/dy$ and $\langle p_{\rm T} \rangle$ from the above sources are listed in Table~\ref{tab_systematic}. 

\begin{table}
\begin{center}
\caption{Systematic uncertainties for the $p_{\rm T}$ spectra, $dN/dy$ and $\langle p_{\rm T} \rangle$ of $K^{*0}$ at $\sqrt{s_{NN}}$ = 7.7 - 39 GeV. }
\label{tab_systematic}
\scalebox{1.0}{
\begin{tabular}{ccccc}
\hline 
\hline
 Systematic uncertainties & spectra & $dN/dy$ & $\langle p_{\rm T} \rangle$  \\
\hline \hline

fitting region &  1-3\% &  1\% &  1\% \\

residual background & 2-4\% &   1-2\% &  1\% \\

fitting function & $\approx$ 1\%  &  $\approx$ 1\% & $\approx$ 1\%  \\


yield extraction &  4\% &  4\% &  1\% \\

particle identification & 2-5\% &  1-2\% &  1-2\% \\

track selection & 1-3\% &  1-2\% & 1-2\% \\

tracking efficiency & 7.1\% &  7.1\% & 7.1\% \\

low $p_{\rm T}$ extrapolation & -- &  5-6\% & 3\% \\
 
 width fix/free & 2-3\% & 2-3\% & 1\% \\
 
Total & 9-12\% & 10-11\% & 8-8.5\% \\

 \hline \hline
\end{tabular}
}
\end{center}

\end{table}

\section{Results}

\begin{figure*}[!ht]
\begin{center}
\includegraphics[scale=0.9]{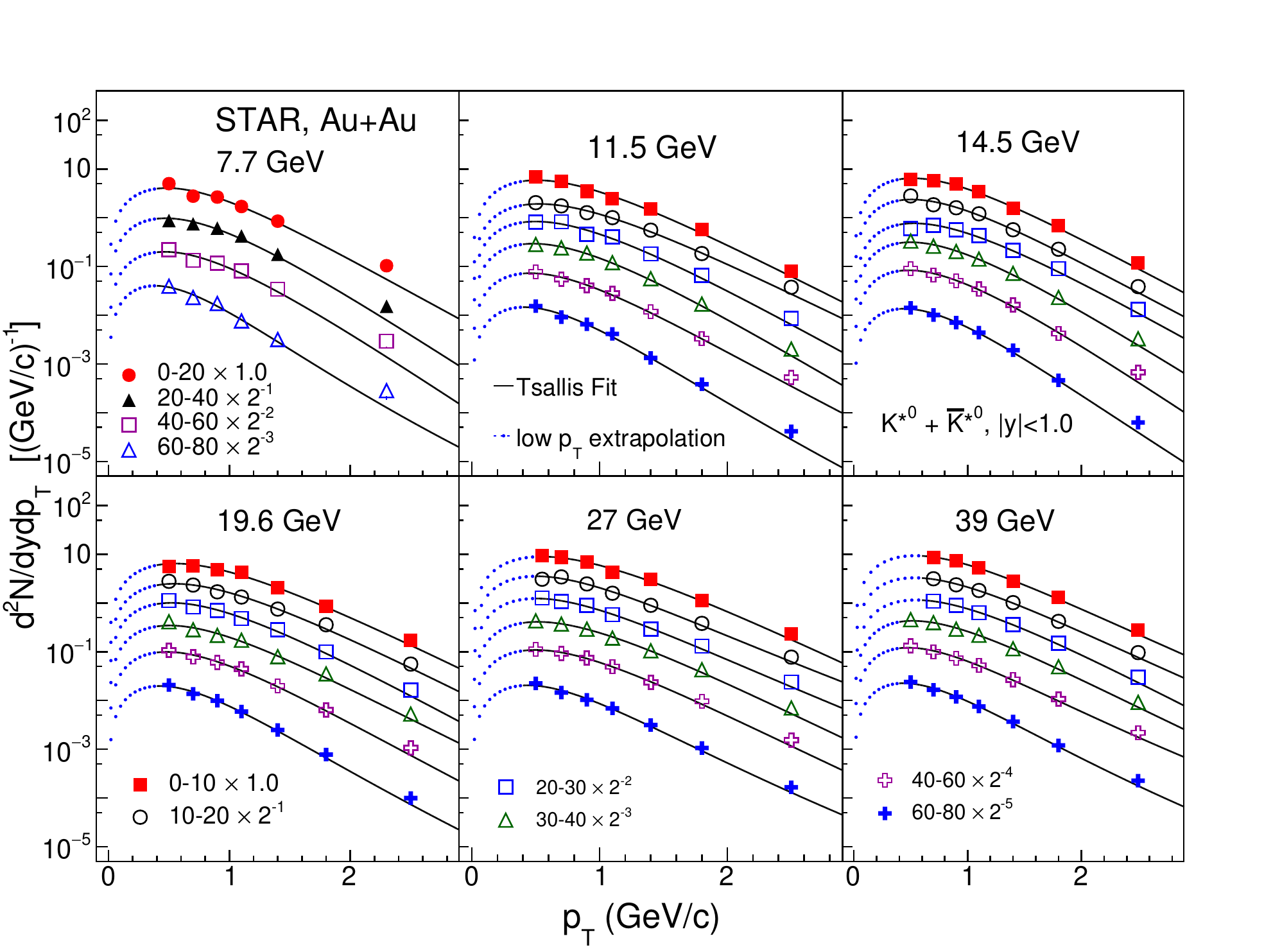}
\caption{$K^{*0}$ transverse momentum ($p_{\rm T}$) spectra at mid-rapidity ($|y| < 1$) for various collision centrality intervals in Au+Au collisions at $\sqrt{s_{\rm NN}}$ = 7.7, 11.5, 14.5, 19.6, 27 and 39 GeV. The solid  and dashed lines indicate the Tsallis fit to the data and its extrapolation to the un-measured low $p_{\rm T}$ region. The statistical and systematic uncertainties are within the marker size.}
\label{fig-pt-spectra-bes}
\end{center}
\end{figure*}

\subsection{Transverse momentum spectra}
The raw yield of $K^{*0}$ is normalized to the number of events ($N_{\mathrm{evt}}$), corrected for detector acceptance $\times$ reconstruction efficiency ($\epsilon_{\mathrm{acc \times rec}}$), particle identification efficiency ($\epsilon_{\mathrm{PID}}$) and branching ratio (BR),
\begin{equation}
\frac{ d^{2}N}{ dp_{\rm T}dy} = \frac{1}{N_{\mathrm{evt}}} \times
\frac{N^{\mathrm{raw}}}{dy dp_{\rm T} } \times \frac{1}{\epsilon_{\mathrm{acc \times rec}} \times \epsilon_{\mathrm{PID}}
  \times \mathrm{BR}},\hspace{-.5em}
 \label{eqn-spectra}
\end{equation}

Figure~\ref{fig-pt-spectra-bes} presents the $K^{*0}$ $p_{\rm T}$ spectra at mid rapidity ($|y| < 1.0$) for various collision centrality intervals in Au+Au collisions at $\sqrt{s_{\rm NN}}$ = 7.7, 11.5, 14.5, 19.6, 27 and 39 GeV. The data are fitted with a Tsallis function~\cite{Tsallis:1987eu} and defined by,

\begin{equation}
\frac{ d^{2}N}{ dp_{\rm T}dy} = p_{\rm T} \frac{(n-1)(n-2)}{nT + (nT+ m(n-2))}  \frac{dN}{dy} ( 1+ \frac{m_{\rm T} - m}{nT})^{n},
\end{equation}
where $m_{\rm T}= \sqrt{m^{2} + p_{\rm T}^{2}}$, T is the inverse slope parameter and $n$ is the exponent. The Tsallis function describes both the exponential shape at low $p_{\rm T}$ and power law at high $p_{\rm T}$. The Tsallis function is found to fit the spectra reasonably well across all the collision centrality intervals and beam energies with $\chi^{2}/\mathrm{NDF} < 2$. The Tsallis fit is used to extrapolate the yield in the un-measured $p_{\rm T}$ regions. The typical range of fit parameters obtained are 12-100 for $n$ and 150-285 MeV for $T$, respectively.

\subsection{Yield and mean transverse momentum}

\begin{figure*}[!ht]
\begin{center}
\includegraphics[scale=0.5]{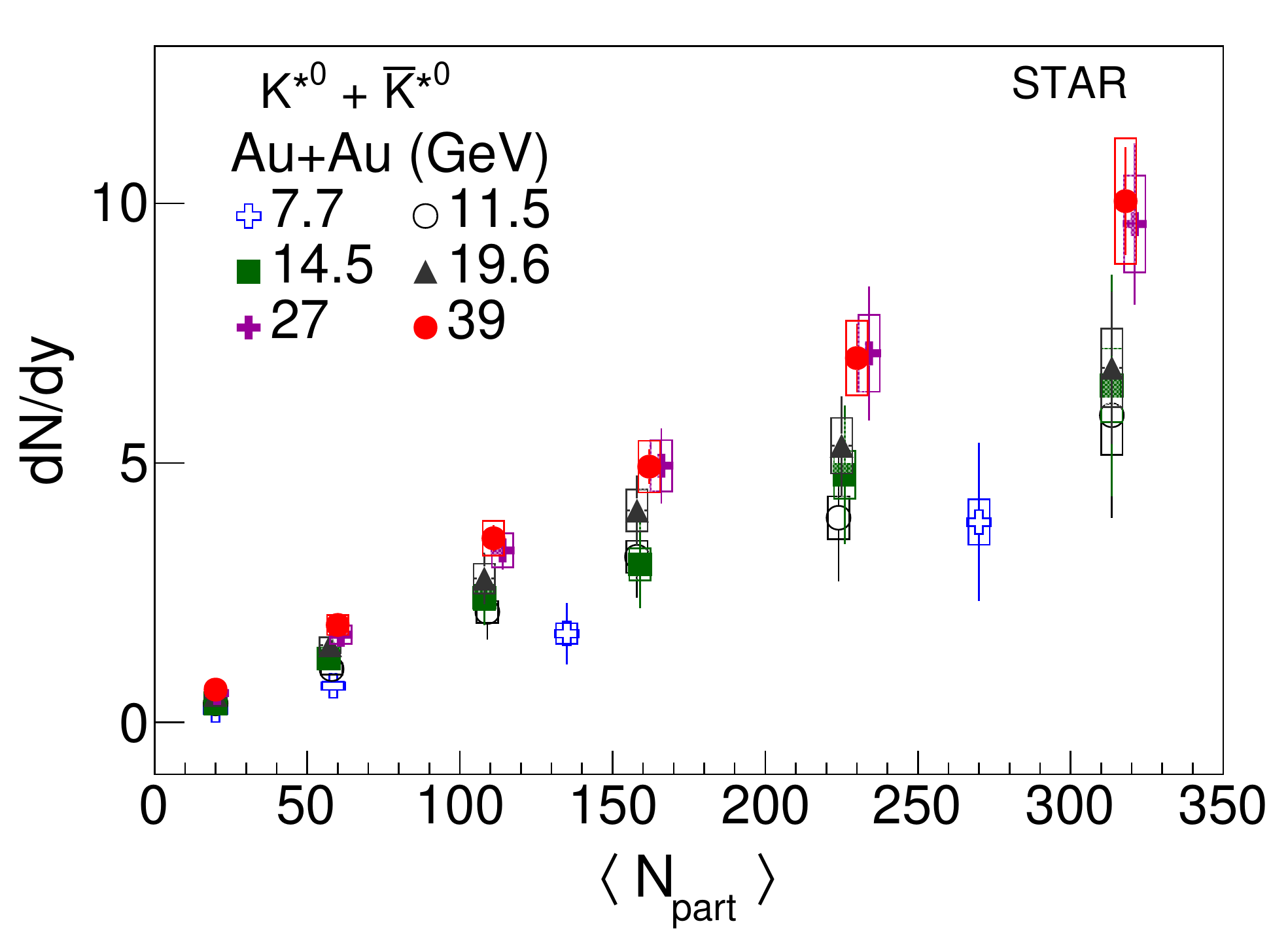}
\caption{ Mid-rapidity yield of $K^{*0}$ as a function of average number of participating nucleons in Au+Au collisions at $\sqrt{s_{\rm NN}}$ = 7.7, 11.5, 14.5, 19.6, 27 and 39 GeV. The vertical bars and open boxes respectively denote the statistical and systematic uncertainties.}
\label{fig-dndy-npart-bes}
\end{center}
\end{figure*}

\begin{figure*}[!ht]
\begin{center}
\includegraphics[scale=0.9]{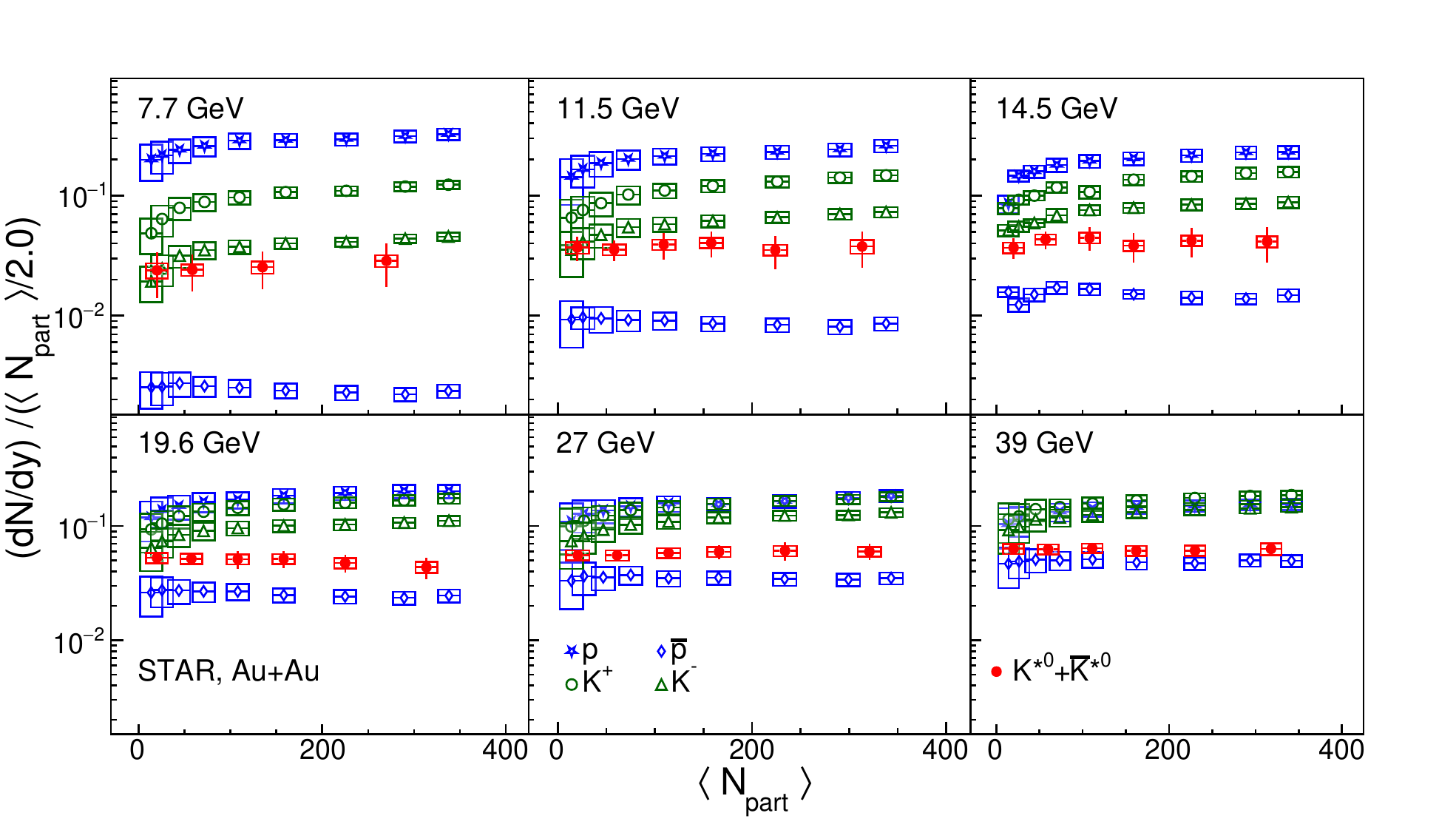}
\caption{ Mid-rapidity yield per average number of participating nucleons  for $K^{*0}$, $K^{\pm}$, p and $\bar{p}$ as a function of $\langle N_{\mathrm{part}} \rangle$ from Au+Au collisions at $\sqrt{s_{\rm NN}}$ = 7.7, 11.5, 14.5, 19.6, 27 and 39 GeV. The vertical bars and open boxes denote the statistical and systematic uncertainties, respectively.}
\label{fig-dndy-npart-scale_bes}
\end{center}
\end{figure*}

The $K^{*0}$ $dN/dy$ is calculated using measured  $p_{\rm T}$ spectra and assuming Tsallis fit function for extrapolation into the un-measured $p_{\rm T}$  region. The low $p_{\rm T}$ extrapolation accounts for 20-40\% of $K^{*0}$ yield. Figure~\ref{fig-dndy-npart-bes} presents the $K^{*0}$ $dN/dy$ as a function of average number of participating nucleons ($\langle N_{\mathrm{part}}\rangle$) in Au+Au collisions at $\sqrt{s_{\rm NN}}$ = 7.7, 11.5, 14.5, 19.6, 27 and 39 GeV. 
The $dN/dy$ is approximately linear with $\langle N_{\mathrm{part}} \rangle$. Figure~\ref{fig-dndy-npart-scale_bes} presents the centrality dependence of $dN/dy$ per average number of participant nucleons for $K^{*0}$. Results are compared with corresponding BES-I measurements of $K^{\pm}$, p and $\bar{p}$\cite{STAR:2017sal,STAR:2019vcp}. On the contrary to $K^{\pm}$ and p, the normalized $K^{*0}$ yield shows a weak dependence on centrality similar to $\bar{p}$.

\begin{figure*}[!ht]
\begin{center}
\includegraphics[scale=0.5]{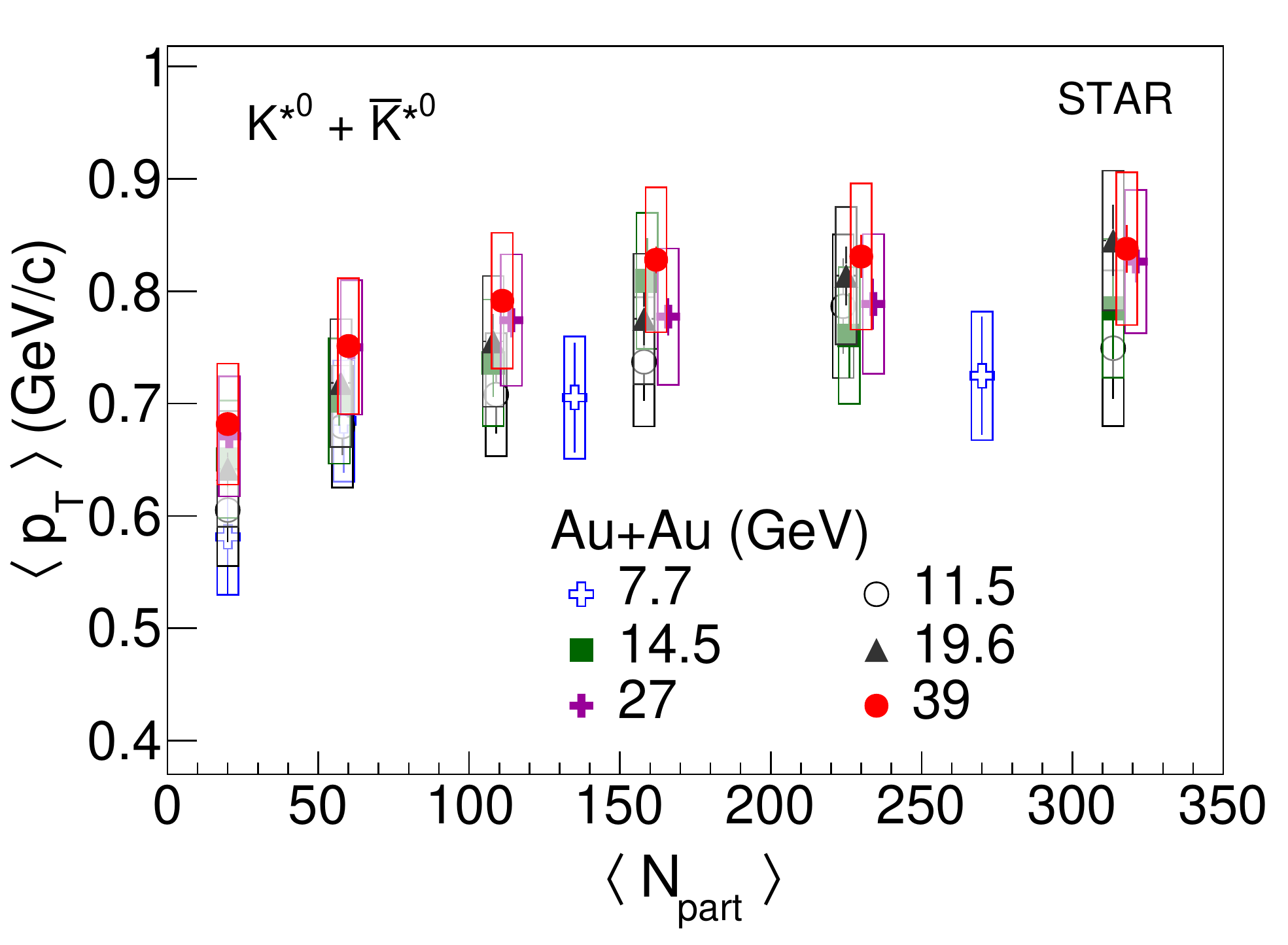}
\caption{The mean transverse momentum of $K^{*0}$ as a function of $\langle N_{\mathrm{part}} \rangle$ in Au+Au collisions at $\sqrt{s_{\rm NN}}$ = 7.7, 11.5, 14.5, 19.6, 27 and 39 GeV. The vertical bars and open boxes respectively denote the statistical and systematic uncertainties.}
\label{fig:kstar-mpt}
\end{center}
\end{figure*}

\begin{figure*}[!ht]
\begin{center}
\includegraphics[scale=0.9]{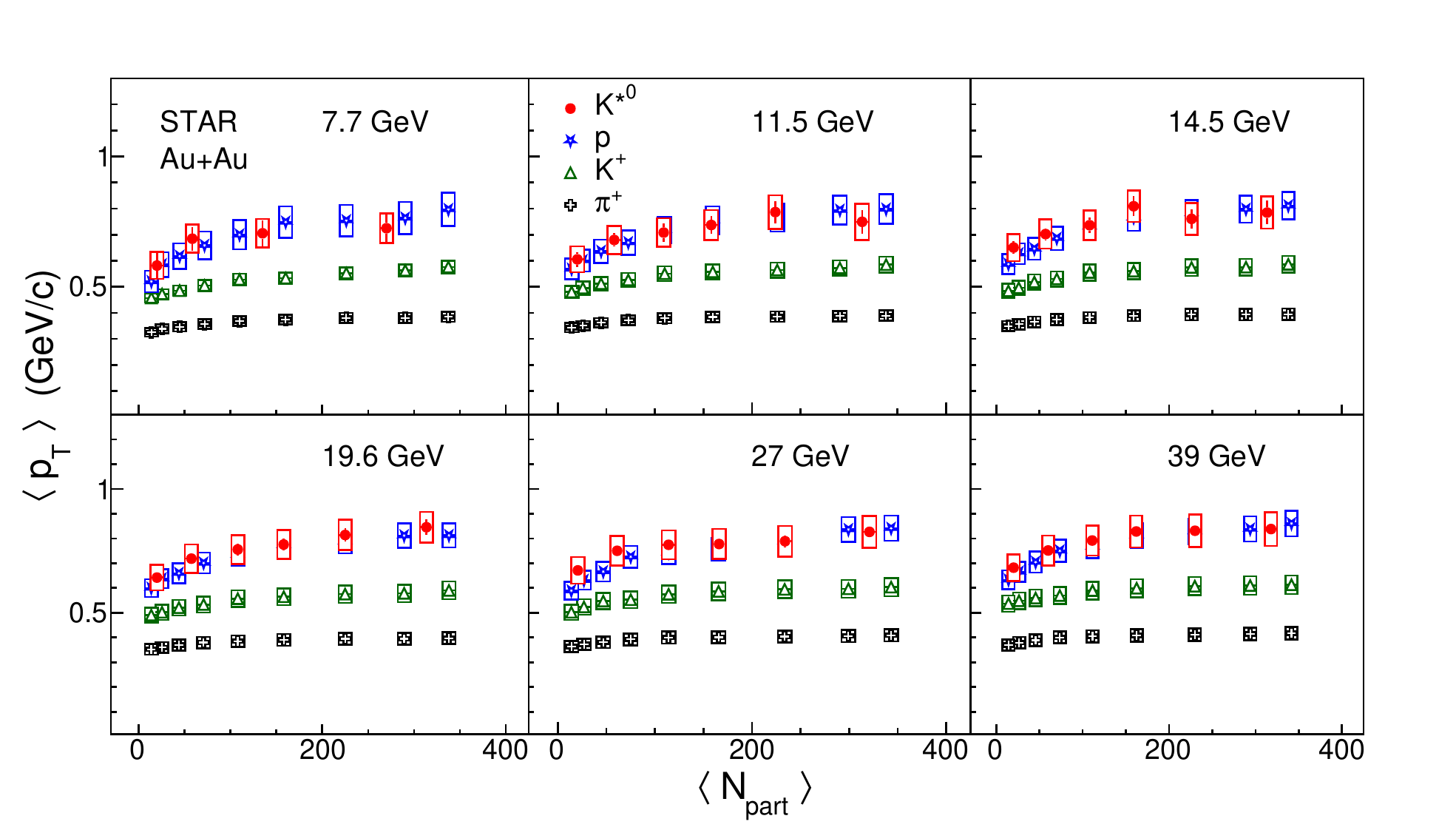}
\caption{The average transverse momentum of $\pi$, K, p~\cite{STAR:2017sal,STAR:2019vcp} and $K^{*0}$ as a function of average number of participating nucleons in Au+Au collisions at $\sqrt{s_{\rm NN}}$ = 7.7, 11.5, 14.5, 19.6, 27 and 39 GeV. The vertical bars and boxes denote the statistical and systematic uncertainties, respectively.}
\label{fig:kstar-mpt-wbes1pid}
\end{center}
\end{figure*}

In Figure~\ref{fig:kstar-mpt}, the $K^{*0}$ $\langle p_{\rm T} \rangle$ is estimated using measured $p_{\rm T}$ spectra and extrapolated to the un-measured $p_{\rm T}$ regions. 
The $K^{*0}$  $\langle p_{\rm T} \rangle$ is also compared with other identified particle species: $\pi$, K, and p as shown in Figure~\ref{fig:kstar-mpt-wbes1pid}.  The $\langle p_{\rm T} \rangle$ of $K^{*0}$ is higher than pions and kaons, and consistent with that of protons~\cite{STAR:2017sal,STAR:2019vcp}. The trend suggests that the $\langle p_{\rm T} \rangle$ is strongly coupled with the mass of the particle and consistent with previous RHIC observations~\cite{STAR:2004bgh,STAR:2008twt,STAR:2010avo}. Considering the systematic uncertainty that is not correlated in centrality bins (i.e. excluding the uncertainty in tracking efficiency $\approx$ 7.1\% which is correlated among all centrality bins), the observed  increase in $\langle p_{\rm T} \rangle$ from peripheral to central collisions is consistent with expectations from increasing radial flow in more central collisions. Moreover, the contributions from hadronic rescattering can also increase $\langle p_{\rm T} \rangle$ in central collisions~\cite{Knospe:2015nva}. 
Table~\ref{tab_dndy_mpt_ratio} presents the $dN/dy$ and $\langle p_{\rm T} \rangle$ of $K^{*0}+\overline{K}^{*0}$ at different collision centrality intervals and beam energies.


\begin{table*}
\caption{$dN/dy$ and $\langle p_{\rm T} \rangle$ of $K^{*0}+\overline{K}^{*0}$, $(K^{*0}+\overline{K}^{*0})/(K^{+} + K^{-})$ ratio at $\sqrt{s_{NN}}$ = 7.7 - 39 GeV. The uncertainties represent statistical and systematic uncertainties, respectively.}
\begin{tabular}{ p{2cm} p{2cm} p{3.5cm} p{3.5cm} p{3.5cm} }
\multicolumn{5}{c}{} \\
$\sqrt{s_{NN}}$ (GeV) & Centrality & \ \ \ \  \ \ $dN/dy$ & \ \ \ \ \ $\langle p_{\rm T} \rangle$ (GeV/c) & \ \ \ \ \ \ $K^{*0}/K$  \\
\hline \hline 
\\
& 0-20\% & 3.86 $\pm$  1.52 $\pm$ 0.43  & 0.725 $\pm$ 0.052 $\pm$ 0.057  &  0.167 $\pm$ 0.066 $\pm$ 0.018\\
& 20-40\% & 1.71 $\pm$  0.59 $\pm$ 0.2  & 0.705 $\pm$ 0.048 $\pm$ 0.054 &  0.178 $\pm$ 0.062 $\pm$ 0.021 \\
7.7  & 40-60\% & 0.70 $\pm$  0.23 $\pm$ 0.07  & 0.684 $\pm$ 0.045 $\pm$ 0.054 &  0.203 $\pm$ 0.068 $\pm$ 0.027 \\
& 60-80\% & 0.24 $\pm$  0.10 $\pm$ 0.04  & 0.581 $\pm$ 0.051 $\pm$ 0.051  &  0.297 $\pm$ 0.122 $\pm$ 0.060\\ \\

\hline \\
& 0-10\% & 5.92 $\pm$  1.98 $\pm$ 0.76 & 0.750 $\pm$ 0.045 $\pm$ 0.069  &  0.173 $\pm$ 0.058 $\pm$ 0.022  \\
& 10-20\% & 3.94 $\pm$  1.22 $\pm$ 0.41 & 0.786 $\pm$ 0.042 $\pm$ 0.063  &  0.177 $\pm$ 0.055 $\pm$ 0.018 \\
11.5  & 20-30\% & 3.19 $\pm$  0.78 $\pm$ 0.30  & 0.737 $\pm$ 0.035 $\pm$ 0.057  &  0.220 $\pm$ 0.054 $\pm$ 0.022 \\
& 30-40\% & 2.13 $\pm$  0.53 $\pm$ 0.21  & 0.707 $\pm$ 0.034 $\pm$ 0.054  &  0.230 $\pm$ 0.058 $\pm$ 0.025 \\
& 40-60\% & 1.03 $\pm$  0.20 $\pm$ 0.10  & 0.679 $\pm$ 0.025 $\pm$ 0.054  &  0.238 $\pm$ 0.046 $\pm$ 0.031 \\
& 60-80\% & 0.37 $\pm$  0.08 $\pm$ 0.04 & 0.605 $\pm$ 0.028 $\pm$ 0.049  &  0.332 $\pm$ 0.075 $\pm$ 0.063 \\ \\

\hline \\
& 0-10\% & 6.49 $\pm$  2.13 $\pm$ 0.70  & 0.784 $\pm$ 0.045 $\pm$ 0.061  &  0.170 $\pm$ 0.056 $\pm$  0.018 \\
& 10-20\% & 4.77 $\pm$  1.34 $\pm$ 0.46 & 0.760 $\pm$ 0.038 $\pm$ 0.060  &  0.184 $\pm$ 0.051 $\pm$  0.018 \\
14.5  & 20-30\% & 3.04 $\pm$  0.84 $\pm$ 0.30  & 0.809 $\pm$ 0.038 $\pm$ 0.060  &  0.178 $\pm$ 0.049 $\pm$ 0.018 \\
& 30-40\% & 2.40 $\pm$  0.53 $\pm$ 0.24  & 0.736 $\pm$ 0.030 $\pm$ 0.056 &  0.220 $\pm$ 0.048 $\pm$ 0.022 \\
& 40-60\% & 1.23 $\pm$  0.20 $\pm$ 0.12  & 0.702 $\pm$ 0.022 $\pm$ 0.055  & 0.246  $\pm$ 0.040 $\pm$ 0.024 \\
& 60-80\% & 0.36 $\pm$  0.07 $\pm$ 0.03  & 0.650 $\pm$ 0.025 $\pm$ 0.052  & 0.261  $\pm$ 0.050 $\pm$  0.026 \\ \\

\hline \\
& 0-10\% & 6.83 $\pm$  1.47 $\pm$ 0.75  & 0.845 $\pm$ 0.031 $\pm$ 0.062  &  0.154 $\pm$ 0.033 $\pm$ 0.017  \\
& 10-20\% & 5.33 $\pm$  0.95 $\pm$ 0.53  & 0.813 $\pm$ 0.026 $\pm$ 0.061  &  0.180 $\pm$ 0.032 $\pm$ 0.018 \\
19.6  & 20-30\% & 4.08 $\pm$  0.67 $\pm$ 0.40 & 0.775 $\pm$ 0.023 $\pm$ 0.058  &  0.201 $\pm$ 0.033 $\pm$ 0.021  \\
& 30-40\% & 2.77 $\pm$  0.50 $\pm$ 0.28  & 0.755 $\pm$ 0.024 $\pm$ 0.058  &  0.213 $\pm$ 0.038 $\pm$ 0.024  \\
& 40-60\% & 1.48 $\pm$  0.16 $\pm$ 0.15 & 0.718 $\pm$ 0.015 $\pm$ 0.057  &  0.238 $\pm$ 0.026 $\pm$ 0.031 \\
& 60-80\% & 0.52 $\pm$  0.06 $\pm$ 0.05 & 0.641 $\pm$ 0.014 $\pm$ 0.051  &  0.312 $\pm$ 0.035 $\pm$  0.056  \\ \\

\hline \\
& 0-10\% & 9.60 $\pm$  1.56 $\pm$ 0.93  & 0.826 $\pm$ 0.018 $\pm$ 0.063 &  0.195 $\pm$ 0.032 $\pm$ 0.018  \\
& 10-20\% & 7.11 $\pm$  1.28 $\pm$ 0.73  & 0.788 $\pm$ 0.022 $\pm$ 0.062  &  0.209 $\pm$ 0.038 $\pm$ 0.021 \\
27  & 20-30\% & 4.95 $\pm$  0.72 $\pm$ 0.49  & 0.777 $\pm$ 0.016 $\pm$ 0.060  &  0.216 $\pm$ 0.031 $\pm$ 0.022 \\
& 30-40\% & 3.31 $\pm$  0.36 $\pm$ 0.32 & 0.774 $\pm$ 0.015 $\pm$ 0.058  &  0.228 $\pm$ 0.025 $\pm$ 0.024 \\
& 40-60\% & 1.69 $\pm$  0.14 $\pm$ 0.18  & 0.750 $\pm$ 0.011 $\pm$ 0.060  &  0.240 $\pm$ 0.020 $\pm$ 0.031 \\
& 60-80\% & 0.57 $\pm$  0.04 $\pm$ 0.06  & 0.670 $\pm$ 0.010 $\pm$ 0.053  &  0.300 $\pm$ 0.023 $\pm$ 0.058 \\ \\

\hline \\
& 0-10\% & 10.04 $\pm$  1.04 $\pm$ 1.21  & 0.837 $\pm$ 0.021 $\pm$ 0.067  & 0.191  $\pm$ 0.020 $\pm$ 0.022    \\
& 10-20\% & 7.02 $\pm$  0.65 $\pm$ 0.71 & 0.830 $\pm$ 0.019 $\pm$ 0.065  &  0.194 $\pm$ 0.018 $\pm$ 0.020  \\
39  & 20-30\% & 4.92 $\pm$  0.33 $\pm$ 0.49 & 0.828 $\pm$ 0.012 $\pm$ 0.064  &  0.202 $\pm$ 0.013 $\pm$ 0.021  \\
& 30-40\% & 3.54 $\pm$  0.25 $\pm$ 0.33  & 0.791 $\pm$ 0.010 $\pm$ 0.060  & 0.225  $\pm$ 0.016 $\pm$ 0.023 \\
& 40-60\% & 1.87 $\pm$  0.09 $\pm$ 0.19  & 0.751 $\pm$ 0.006 $\pm$ 0.060 &  0.241 $\pm$ 0.012 $\pm$ 0.031  \\
& 60-80\% & 0.63 $\pm$  0.03 $\pm$ 0.06 & 0.681 $\pm$ 0.006 $\pm$ 0.053 &  0.290 $\pm$ 0.015 $\pm$ 0.052 
\\

 \hline \hline
\end{tabular}

\label{tab_dndy_mpt_ratio}
\end{table*}

\subsection{Particle ratios}

\begin{figure*}[!ht]
\begin{center}
\includegraphics[scale=0.5]{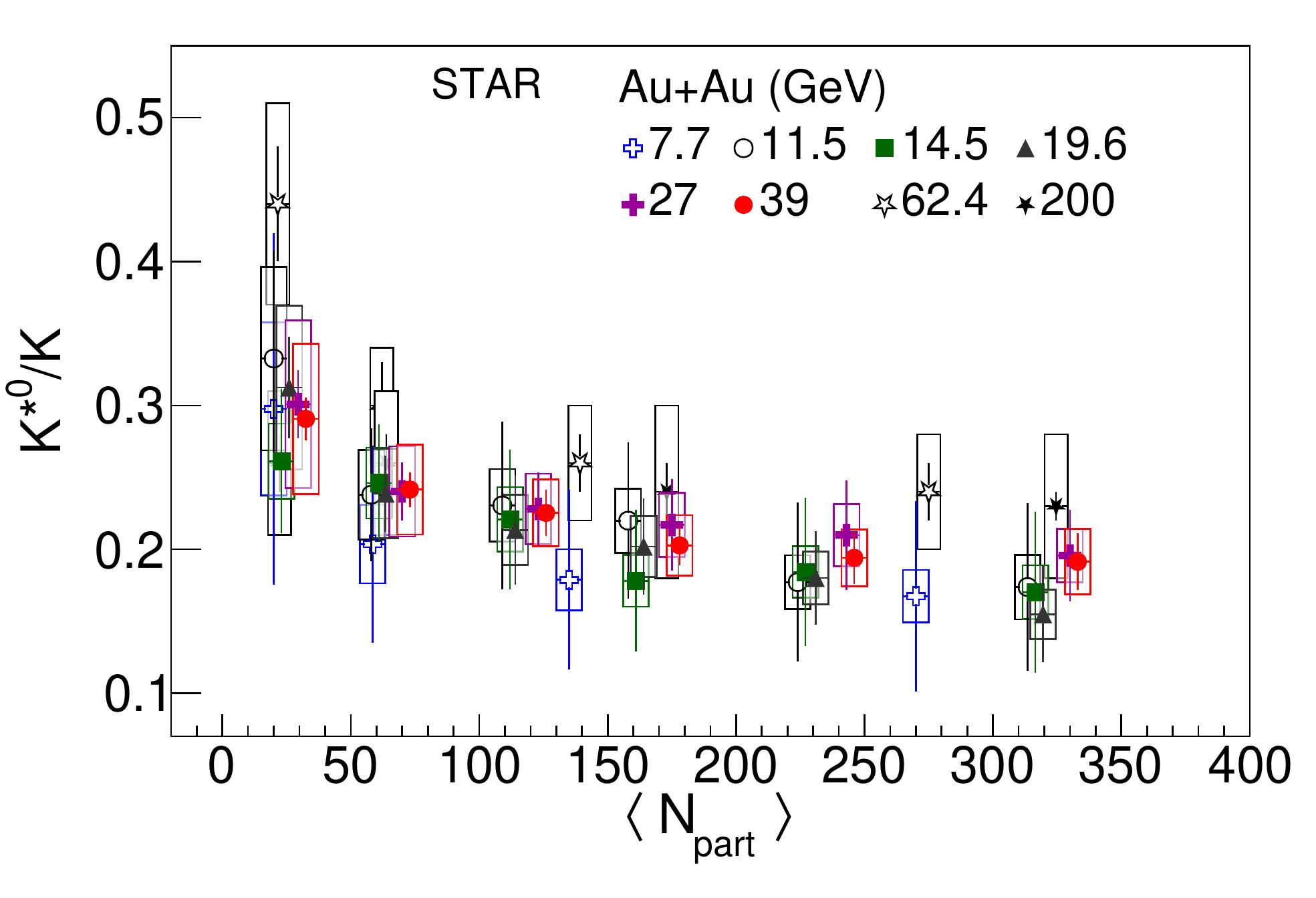}
\caption{ $K^{*0}/K$ ratio at mid rapidity as a function of average number of participating nucleons in Au+Au collisions at $\sqrt{s_{\rm NN}}$ = 7.7, 11.5, 14.5, 19.6, 27 and 39 GeV. The vertical bars and open boxes respectively denote the statistical and systematic uncertainties. The results are compared with previously published STAR~\cite{STAR:2004bgh,STAR:2010avo} measurements.}
\label{fig:kstar2k-npart}
\end{center}
\end{figure*}

\begin{figure*}[!ht]
\begin{center}
\includegraphics[scale=0.5]{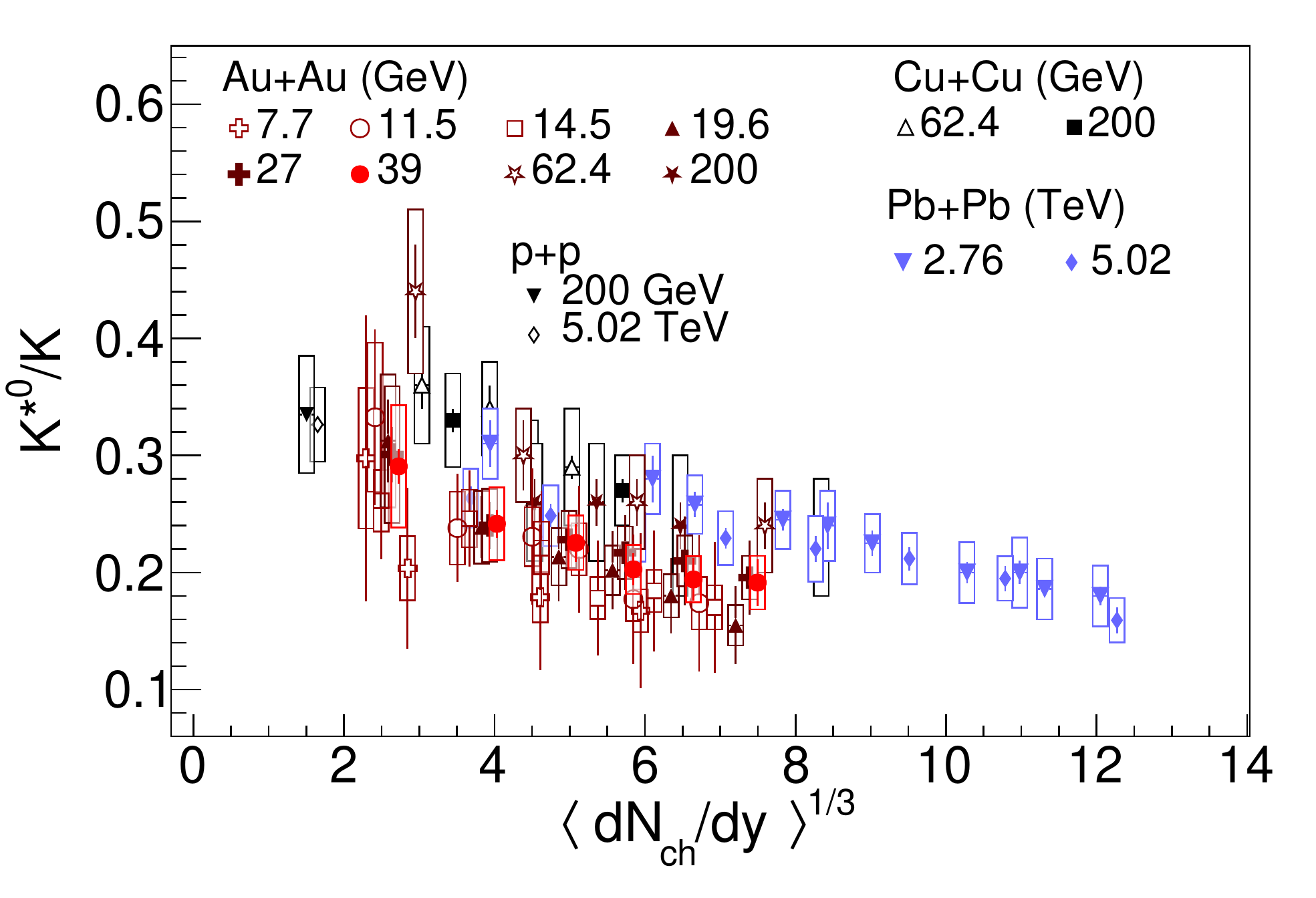}

\caption{ $K^{*0}/K$ ratio at mid rapidity as a function of $\langle dN_{ch}/dy \rangle^{1/3}$ in Au+Au collisions at $\sqrt{s_{\rm NN}}$ = 7.7, 11.5, 14.5, 19.6, 27 and 39 GeV. The vertical bars and open boxes respectively denote the statistical and systematic uncertainties. The results are compared with previously published STAR~\cite{STAR:2004bgh,STAR:2010avo} and ALICE~\cite{ALICE:2014jbq,ALICE:2017ban,ALICE:2019xyr,ALICE:2021ptz} measurements.}
\label{fig:kstar2k-dndy13}
\end{center}
\end{figure*}

\begin{figure*}[!ht]
\begin{center}
\includegraphics[scale=0.5]{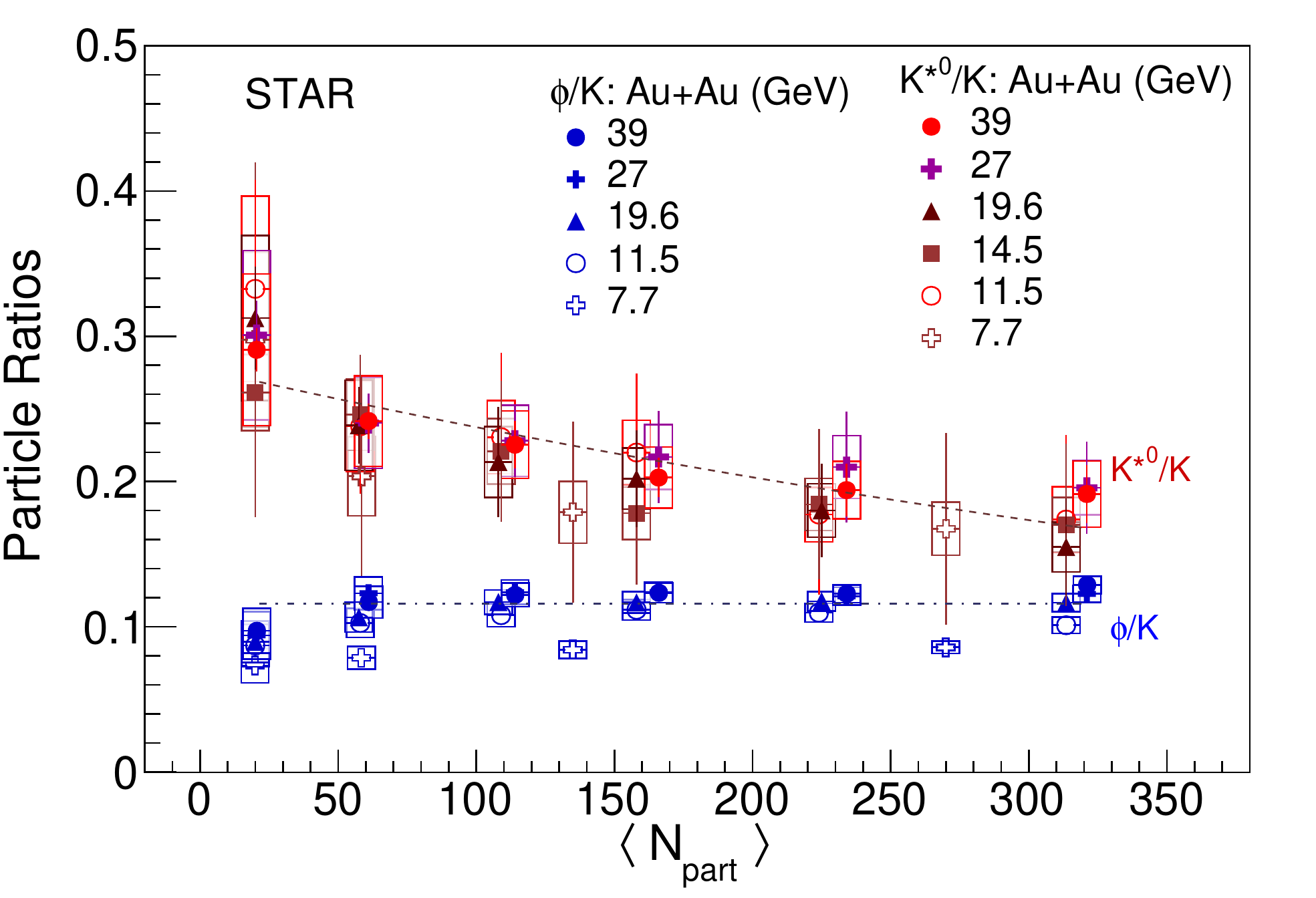}

\caption{Comparison of $K^{*0}/K$ and $\phi/K$~\cite{STAR:2019bjj} ratio at mid rapidity as a function of average number of participating nucleons in Au+Au collisions at $\sqrt{s_{\rm NN}}$ = 7.7 -- 39 GeV. The vertical bars and boxes respectively denote the statistical and systematic uncertainties. The dashed lines are used to guide the eyes.}
\label{fig:kstar2k-phi2k-npart}
\end{center}
\end{figure*}

The ratios of resonances ($K^{*0}$ and $\phi$ ) to the non-resonances have been studied previously in small system (e+e, p+p, p+A and d+A) and heavy-ion (A+A) collisions. Such ratios are useful in understanding the late stage interactions in heavy ion collisions. Since the lifetime of $K^{*0}$ and $\phi$ differ by about a factor of ten, their production can shed light on the different time scales of the evolution of the system in HIC. It is observed by the STAR, ALICE and NA49 experiments that the $K^{*0}/K$ ratio is smaller in central collisions than in peripheral (and small system) collisions. While the $\phi/K$ ratio is observed to be independent of centrality, which is expected due to the longer lifetime of $\phi$ mesons. Figure~\ref{fig:kstar2k-npart} presents the $K^{*0}/K$ ($=(K^{*0}+\overline{K}^{*0})/(K^{+} + K^{-}$)) ratio as a function of $\langle N_{\mathrm{part}} \rangle$ for six different beam energies. The charged kaon yields are taken from~\cite{STAR:2017sal,STAR:2019vcp}. The BES-I results are compared with previously published STAR measurements in Au+Au collisions at $\sqrt{s_{\rm NN}}$ = 62.4 and 200 GeV~\cite{STAR:2004bgh,STAR:2010avo}. The BES-I measurements follow the same centrality dependence as observed in previous measurements. From HBT studies, the variable $\langle dN_{ch}/dy \rangle^{1/3}$ can be considered as a proxy for the system radius in heavy ion collisions. If one assumes that the strength of rescattering is related to the distance travelled by the resonance decay products in the hadronic medium, then one naively expects $K^{*0}/K$ ratio to decrease exponentially with $\langle dN_{ch}/dy \rangle^{1/3}$~\cite{ALICE:2014jbq}. Figure~\ref{fig:kstar2k-dndy13} presents the $K^{*0}/K$ ratio as a function of $\langle dN_{ch}/dy \rangle^{1/3}$ for BES-I energies. These results are compared to previous measurements of different collision systems and beam energies from RHIC~\cite{STAR:2004bgh,STAR:2010avo} and LHC\cite{ALICE:2014jbq,ALICE:2017ban,ALICE:2019xyr,ALICE:2021ptz}. Although present uncertainties in the data preclude any strong conclusion, we observe that the $K^{*0}/K$ ratios from all BES energies follow the same behavior and those from LHC energies seem to be slightly larger. 
Figure~\ref{fig:kstar2k-phi2k-npart} compares the $K^{*0}/K$ and $\phi/K (=2\phi/(K^{+}+K^{-}))$~\cite{STAR:2019bjj} ratios in Au+Au collisions at $\sqrt{s_{\rm NN}}$ = 7.7 -- 39~GeV. Unlike $K^{*0}/K$, the $\phi/K$ ratio is mostly observed to be independent of collision centrality at these energies. The centrality dependent trend of $K^{*0}/K$ and $\phi/K$ ratio is consistent with the expectation of more rescattering in more central collisions for $K^{*0}$ daughters. 

\begin{figure*}[!ht]
\begin{center}
\includegraphics[scale=0.5]{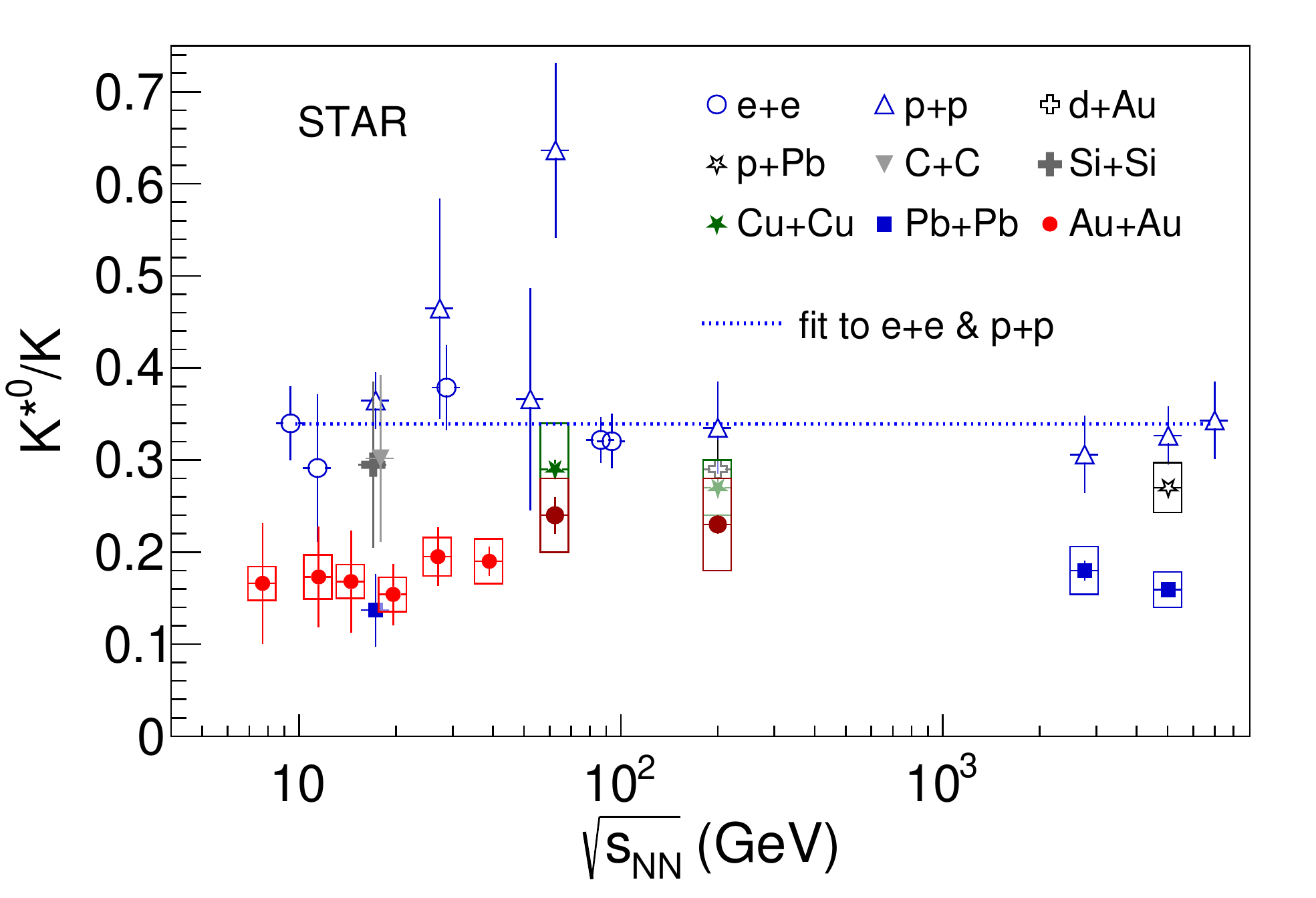}
\caption{The beam energy dependence of $K^{*0}/K$ ratio in e+e~\cite{ARGUS:1993ggm,Pei:1996kq,Hofmann:1988gy,SLD:1998coh}, p+p~\cite{Aguilar-Benitez:1991hzq,STAR:2004bgh,AnnecyLAPP-CERN-CollegedeFrance-Dortmund-Heidelberg-Warsaw:1981whv,AxialFieldSpectrometer:1982btk}, d+Au~\cite{STAR:2008twt}, p+Au~\cite{ALICE:2016sak,ALICE:2021uyz} and most-central C+C, Si+Si~\cite{NA49:2011bfu}, Au+Au~\cite{STAR:2004bgh,STAR:2010avo} and Pb+Pb~\cite{ALICE:2014jbq,ALICE:2017ban,ALICE:2019xyr} collisions. For e+e and p+p collisions, the bars denote the quadratic sum of statistical and systematic uncertainties. For p+A and A+A data, the bars denote the statistical uncertainties and the boxes denote the systematic uncertainties.}
\label{fig-kstar2k-snn}
\end{center}
\end{figure*}

The measurement of $K^{*0}/K$ ratio in a broad beam energy range may provide information on production mechanisms, especially the energy dependence of the relative strength of rescattering and regeneration processes. Figure~\ref{fig-kstar2k-snn} presents the beam energy dependence of $K^{*0}/K$ ratio in small systems (e+e~\cite{ARGUS:1993ggm,Pei:1996kq,Hofmann:1988gy,SLD:1998coh}, p+p~\cite{Aguilar-Benitez:1991hzq,STAR:2004bgh,AnnecyLAPP-CERN-CollegedeFrance-Dortmund-Heidelberg-Warsaw:1981whv,AxialFieldSpectrometer:1982btk}, d+Au~\cite{STAR:2008twt} and p+Pb~\cite{ALICE:2016sak,ALICE:2021uyz}) and in central heavy-ion (C+C, Si+Si, Au+Au and Pb+Pb~\cite{NA49:2011bfu,STAR:2004bgh,STAR:2010avo,ALICE:2014jbq,ALICE:2017ban,ALICE:2019xyr}) collisions. The $K^{*0}/K$ ratio is independent of beam energy in small system collisions. The data, with combined statistical and systematic uncertainties, is fitted to a straight line and the resulting value is 0.34 $\pm$ 0.01. The $K^{*0}/K$ from  STAR BES-I energy is found to be consistent with that from Pb+Pb collisions at $\sqrt{s_{\rm NN}}$=17.3 GeV by NA49~\cite{NA49:2011bfu}. Overall, there is a suppression of $K^{*0}/K$ ratio in central heavy-ion collisions relative to the small system collisions.  The smaller $K^{*0}/K$ ratio in heavy-ion collisions compared to small system collisions is consistent with the expectation from the dominance of rescattering over regeneration in most-central heavy-ion collisions. 


\begin{figure*}[!ht]
\begin{center}
\includegraphics[scale=0.5]{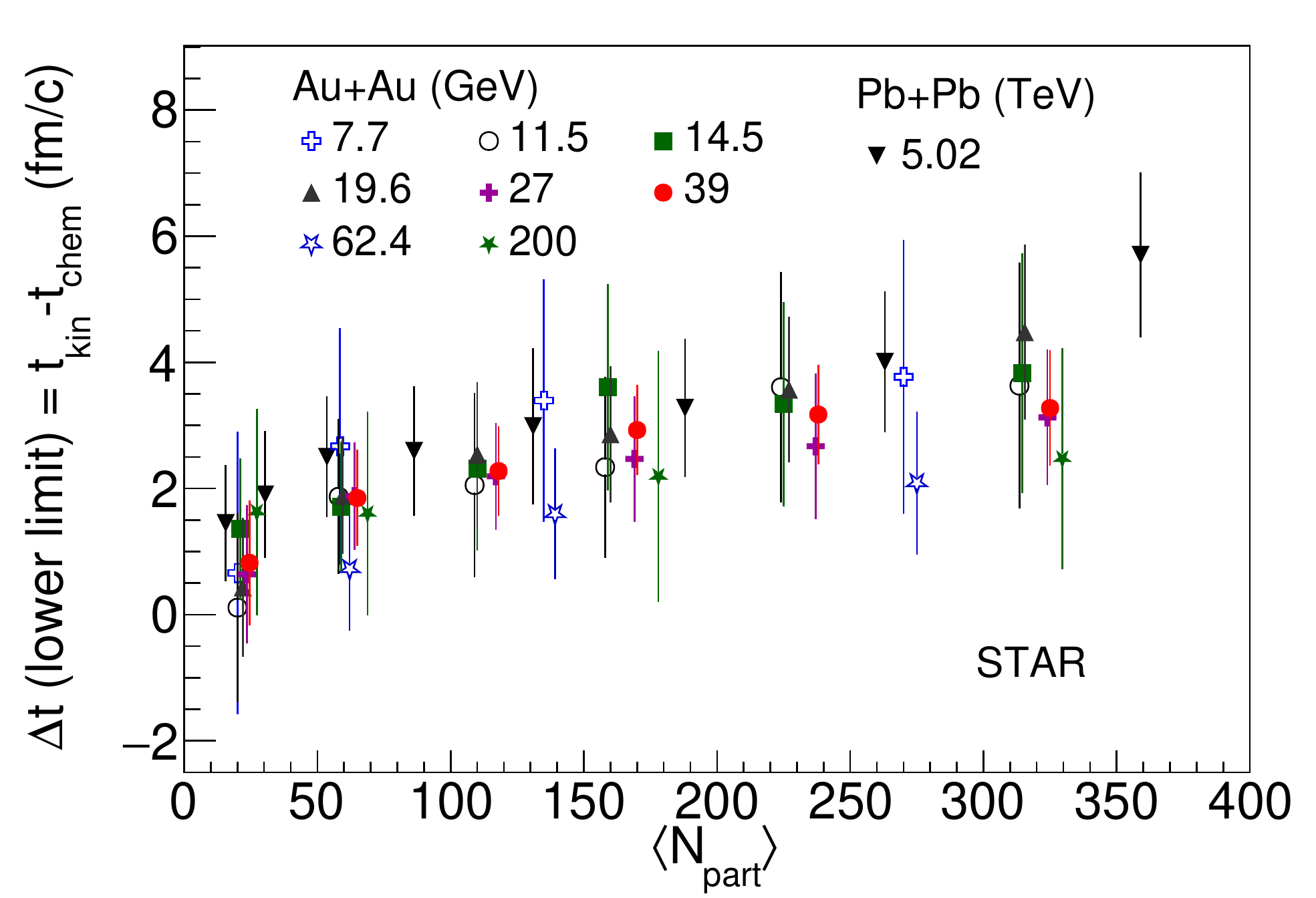}
\caption{The lower limit on the time difference ($\Delta t$) between the chemical and kinetic freeze-out as a function of average number of participating nucleons ($\langle N_{\mathrm{part}} \rangle$). The results are compared with previous STAR~\cite{STAR:2004bgh,STAR:2010avo} and ALICE~\cite{ALICE:2014jbq,ALICE:2017ban,ALICE:2019xyr} measurements.The bars denote combined statistical and systematic uncertainties which is propagated from the uncertainties in $K^{*0}/K$ ratio.}
\label{fig-lifetime-bes}
\end{center}
\end{figure*}

Due to the dominance of rescattering over regeneration, the reaction $K^{*0} \leftrightarrow K\pi$ may not be in balance. Experimentally we can not measure the particle yield ratios at different freeze-outs. Thus we make the approximation that the $(K^{*0}/{K})_{\mathrm{CFO}}$ and $(K^{*0}/{K})_{\mathrm{KFO}}$ are the same as the $K^{*0}/{K}$ ratio measured in elementary and heavy-ion collisions respectively. Furthermore, we assume that (i) all $K^{*0}$ decayed before kinetic freeze-out are lost due to rescattering and (ii) no $K^{*0}$ regeneration occurs between the chemical and kinetic freeze out. Under these assumptions, the $K^{*0}/K$ ratio at different freeze-outs are related in the following way~\cite{STAR:2002npn}, 
\begin{equation}
\bigg(\frac{K^{*0}}{K}\bigg)_{\mathrm{KFO}} = \bigg(\frac{K^{*0}}{K}\bigg)_{\mathrm{CFO}} \times e^{-\Delta t /\tau_{K^{*0}}}, \hspace{-.5em} 
\end{equation}
where $\tau_{K^{*0}}$ is the lifetime of $K^{*0}$ ($\approx$ 4.16 fm/c) and $\Delta t$ is the lower limit of the time difference between CFO and KFO. It has been shown by AMPT calculations that such assumptions are applicable~\cite{Singha:2015fia}. Due to the unavailability of small system collisions at BES-I energies, the $(K^{*0}/{K})_{\mathrm{CFO}}$ is taken from the straight line fit through the global small system data (e+e and p+p data shown in Fig~\ref{fig-kstar2k-snn}). The $(K^{*0}/{K})_{\mathrm{KFO}}$ values are taken from the $K^{*0}/{K}$ measurements at BES-I energies. The estimated $\Delta t$ is boosted by the Lorentz factor~\cite{Singha:2015fia}. Figure~\ref{fig-lifetime-bes} presents the lower limit of the time difference between chemical and kinetic freeze-out as a function of $\langle N_{\mathrm{part}} \rangle$. The $\Delta t$ from BES-I energies are compared with the results from Au+Au collisions at 62.4 and 200 GeV~\cite{STAR:2004bgh,STAR:2010avo}, and Pb+Pb collisions at 5.02 TeV~\cite{ALICE:2019xyr}. The $\Delta t$ from BES-I seems to follow the trend observed in previous RHIC and LHC data. Present uncertainty in BES-I data does not allow determination of the energy dependence of $\Delta t$. Future high-statistics BES-II measurements will offer better precision.

\section{Conclusion}
In summary, we presented the $p_{\rm T}$ spectra, $dN/dy$, and $\langle p_{\rm T}\rangle$ of $K^{*0}$ at mid-rapidity in Au+Au collisions at $\sqrt{s_{\rm NN}}$  = 7.7 - 39 GeV using the $1^{st}$ phase of RHIC beam energy scan data. For BES-I energies, the $K^{*0}$ $\langle p_{\rm T} \rangle$ is larger than that of pions and kaons and comparable to that of protons, indicating a mass dependence of $\langle p_{\rm T} \rangle$. The $K^{*0}/K$ ratio in the most-central Au+Au collisions is smaller than the same in small system collision data. The $K^{*0}/K$ ratio shows a weak centrality dependence and follows the same trend observed by previous RHIC and LHC measurements. On the contrary, the $\phi/K$ ratio is mostly independent of centrality. These observations support the scenario of the dominance of hadronic rescattering over regeneration for $K^{*0}$ at BES energies. Based on the $K^{*0}/K$ ratio, the lower limit of the time between chemical and kinetic freeze-out at BES energies is estimated. The high statistics data from the 2$^{nd}$ phase of BES (BES-II) will allow more precise measurements of hadronic resonances at these energies.

\noindent{\bf Acknowledgments}\\

We thank the RHIC Operations Group and RCF at BNL, the NERSC Center at LBNL, and the Open Science Grid consortium for providing resources and support.  This work was supported in part by the Office of Nuclear Physics within the U.S. DOE Office of Science, the U.S. National Science Foundation, National Natural Science Foundation of China, Chinese Academy of Science, the Ministry of Science and Technology of China and the Chinese Ministry of Education, the Higher Education Sprout Project by Ministry of Education at NCKU, the National Research Foundation of Korea, Czech Science Foundation and Ministry of Education, Youth and Sports of the Czech Republic, Hungarian National Research, Development and Innovation Office, New National Excellency Programme of the Hungarian Ministry of Human Capacities, Department of Atomic Energy and Department of Science and Technology of the Government of India, the National Science Centre and WUT ID-UB of Poland, the Ministry of Science, Education and Sports of the Republic of Croatia, German Bundesministerium f\"ur Bildung, Wissenschaft, Forschung and Technologie (BMBF), Helmholtz Association, Ministry of Education, Culture, Sports, Science, and Technology (MEXT) and Japan Society for the Promotion of Science (JSPS).

\bibliographystyle{unsrtnat}
\bibliography{main}

\end{document}